\documentclass[prd,superscriptaddress,nofootinbib,amsmath,amssymb,aps,11pt]{revtex4-2}
\usepackage[utf8]{inputenc}

\usepackage{bm}
\usepackage{amsfonts}
\usepackage{latexsym}
\usepackage{graphicx}
\usepackage{amsmath}
\usepackage{anyfontsize}
\usepackage{palatino}
\usepackage{mathpazo}
\usepackage{bigints}
\usepackage{hyperref}
\usepackage{booktabs}
\usepackage[normalem]{ulem}
\usepackage{soul}
\usepackage{comment}

\newcommand{\Mp}{M_{\rm Pl}}
\newcommand{\Veff}{V_{\rm eff}}

\newcommand{\Qtwo}{\ensuremath{\mathcal{Q}_{2}}}
\newcommand{\phitwo}{\ensuremath{\phi_{2}}}
\newcommand{\rhos}{\ensuremath{\rho_{\rm s}}}
\newcommand{\rs}{\ensuremath{r_{\rm s}}}
\newcommand{\Cs}{\ensuremath{C_{\rm s}}}

\newcommand{\phiinf}{\ensuremath{\phi_{\infty}}}

\usepackage{xcolor}     

\begin{document}
\title{\boldmath The shape of the Chameleon fifth-force on the mass components of galaxy clusters}

\author{Lorenzo Pizzuti}
\email{lorenzo.pizzuti@unimib.it}
\affiliation{Dipartimento di Fisica G. Occhialini, Universit\'a degli Studi di Milano Bicocca, Piazza della Scienza 3, I-20126 Milano, Italy}

\author{Valentina Amatori}
\affiliation{Department of Physics, ETH Zurich, Otto-Stern-Weg 1, 8093 Zurich, Switzerland}

\author{Alexandre M. Pombo}
\affiliation{CEICO, Institute of Physics of the Czech Academy of Sciences, Na Slovance 2, 182 21 Praha 8, Czechia}

\author{Sandeep Haridasu}
\affiliation{SISSA, Via Bonomea 265, 34136 Trieste, Italy\\
Institute for Fundamental Physics of the Universe (IFPU), Via Beirut 2, 34014 Trieste, Italy\\
INFN-Sezione di Trieste, via Valerio 2, 34127 Trieste, Italy}

\begin{abstract}
    In the context of Chameleon gravity, we present a semi-analytical solution of the chameleon field profile in accurately modelled galaxy cluster's mass components, namely: the stellar mass of the Brightest Cluster Galaxy (BCG), the baryonic mass in galaxies (other than the BCG), the mass of the Intra-Cluster Medium (ICM) and the diffuse cold dark matter (CDM). The obtained semi-analytic profile is validated against the numerical solution of the chameleon field equation and implemented in the \textsc{MG-MAMPOSSt} code for kinematic analyses of galaxy clusters in modified gravity scenarios. By means of mock halos, simulated both in GR and in modified gravity, we show that the combination of velocities and positions of cluster member galaxies, along with data of the stellar velocity dispersion profile of the BCG, can impose constraints on the parameter space of the Chameleon model; for a cluster generated in GR, these constraints are at the same level as a joint lensing+kinematics analysis of a cluster modelled with a single mass profile, without the BCG data.
\end{abstract}

\maketitle
%
 \section{Introduction}
%
    The discovery of the late-time accelerated expansion of the Universe~\cite{Riess,teleparallelgravity} presented one of the most formidable mysteries in modern physics. In order to explain the observed acceleration, the cosmological constant~\cite{Weinberg,BeyondlambdaCDM}, $\Lambda$, was incorporated in General Relativity (GR), forming the foundation of the current framework of cosmological research: the standard $\Lambda$CDM model~\cite{lambdaCDM}.
    
    Supported by a broad array of observations~ \cite{observations1,observations2}, the $\Lambda$CDM model is widely regarded as the best descriptor of the observed Universe's expansion history. Nonetheless, despite its successes, the cosmological constant on which the model is based still lacks a fundamental explanation within the framework of standard physics~\cite{TheFADE}.

    In pursuit of a physical motivation for the cosmological constant, several alternatives have been proposed over the past decades. Of particular interest are models that modify GR -- the foundation of the Standard Cosmological model on large scales~\cite{MG,MG2} -- by introducing an additional scalar field (scalar-tensor theories~\cite{ST,ST2}) capable of replicating the cosmological constant's effect~\cite{Hu}. The introduction of this scalar field adds a component to the gravitational force~\cite{Khoury, Burrage_2018}, leaving measurable signatures on the formation and evolution of cosmic structures~\cite{Brax,Brax_2013}.

    On the small-scale, high-density -- when compared with the background density -- Solar System scale, GR's precise validation~\cite{GR} requires any Modified Gravity (MG) model to contain a screening mechanism in order to reconcile with the well-established GR measurements~\cite{Khoury_2010} at small scales while still impacting the dynamics at large, low-density, cosmological scales.

    While several scalar-tensor models have been ruled out by gravitational waves observations (see \textit{e.g.}~\cite{ Kobayashi:2018xvr, Amendola:2017orw}) or highly constrained by observations (\textit{e.g.}~\cite{Kimura:2011dc,Bayarsaikhan_2020,Bellini_2016,Saltas:2019ius,Hiramatsu_2022}), the relative simplicity and the rich phenomenology make them widely studied in different environment. One particularly interesting class of scalar-tensor theories viable at cosmological scales is the Chameleon model~\cite{Chameleon}. In this framework, gravity's effects vary with the environment due to a coupling between the additional scalar field (chameleon field) and matter. As a result, in the presence of the non-trivial scalar field, matter "feels" an additional fifth-force changing the formation and evolution of structures in the Universe.    
    
    In turn, the coupling feedback causes the scalar field to depend on the local energy density, leading to a very large field mass in high-density regions (suppressing the interaction) while containing a small, non-zero mass in low-density regions. The modulation of the scalar field by the environment results in a built-in screening mechanism where the fifth-force is suppressed at small scales (recovering GR) while impacting large-scale dynamics.

    In Chameleon gravity models, the scalar field is characterized by the coupling constant ($Q$), which determines its interaction with matter, and the asymptotic value of the field at infinity ($\phi_{\infty}$). Both parameters have been strongly constrained by high-accuracy observational data across laboratory (\textit{e.g.} \cite{lablimits1,lablimits2}), astrophysical (\textit{e.g.} \cite{astrolimits1,astrolimits2,Benisty2023}), and cosmological scales (\textit{e.g.} \cite{Wilcox:2015kna, Cataneo:2016iav, Pizzuti17, Tamosiunas_2022}).

    Although the parameter space for viable Chameleon theories has been tightly constrained, it is nevertheless interesting to study its phenomenology at cluster scales, where some flexibility remains. Moreover, the peculiar screening mechanism provides an excellent toy model for exploring the possible effects of the fifth force on the mass components of clusters. In the latter, constraints have been usually derived by modelling the total mass distribution of the cluster (baryonic and dark matter) using the Navarro-Frenk-White (NFW) profile~\cite{Navarro:1995iw}. However, recent studies show some clusters favouring an Isothermal or a Hernquist mass profile~\cite{Ettori:2018tus,Boumechta:2023qhd}. This preference for different mass models can yield strong effects on the chameleon field profile -- and the resulting fifth-force. 

    In a previous work~\cite{Pizzuti2024b}, we proposed the study of the Chameleon mechanism by means of a semi-analytical approximation of the chameleon field. This procedure allowed us to investigate the effect of the chameleon mechanism when changing the parametrization of the total cluster's mass profile, showing how the efficiency of the screening -- and, therefore, the constraints obtainable at the cluster's scales -- strongly depend on the assumed mass model. In this paper, instead of considering a single mass profile to describe the entire cluster's mass distribution, we extend the analysis performed in \cite{Pizzuti2024b} by introducing a semi-analytical approximation of the chameleon field profile in an \textit{explicitly decomposed galaxy cluster's key mass components}: the stellar mass of the Brightest Cluster Galaxy (BCG), the baryonic mass in galaxies, the Intra-Cluster Medium (ICM), and the diffuse, cold dark matter (CDM) component. 

    The semi-analytical approximation, based on the method proposed in \cite{Terukina:2013eqa}, offers a straightforward and computationally efficient (when compared with a full numerical computation) strategy for studying the screening mechanism. Besides, it serves as a valuable tool for exploring the relationship between the parameters of the mass profiles and the chameleon field, emphasizing the key physical aspects of the screening mechanism. To ensure the validity of our semi-analytical approach, we compare it with a full numerical solution and confirm that it accurately reproduces the behaviour of the fifth-force with a maximum relative difference of $10^{-2}$. 

    The validated semi-analytical solution is then fed into MG-MAMPOSSt code for kinematic analyses of galaxy clusters in modified gravity frameworks. Using mock galaxy cluster halos simulated in both GR and Chameleon gravity, we demonstrate that combining the velocity and positional data of cluster member galaxies with the stellar velocity dispersion profile of the BCG strongly constrains the Chameleon model's parameter space.

    The paper is structured as follows: Section~\ref{theory} provides an overview of the fundamental theory of Chameleon gravity and the screening mechanism. In Section~\ref{sec:generalscreening}, the semi-analytical expressions for the chameleon field and its gradient are obtained, assuming a four-components model for a cluster matter distribution; Section~\ref{sec:numerical} further compares and validates the semi-analytical approach with full numerical solutions of the (spherical) field equation.
 
%
 \section{Chameleon gravity and field profile} \label{theory}
%
    The Lagrangian of a chameleon (real) scalar field \cite{Khoury:2003aq}, $\phi$, conformally coupled with the matter fields $\psi^{(i)}$ can be written as 
    \begin{equation} \label{eq:lagrangian}
     \mathcal{L}=\sqrt{-g}\left[ -\frac{\Mp R}{2}+\frac{(\partial\phi)^{2}}{2}+V(\phi)\right] +L_{\rm m}\big( \psi^{(i)},g_{\mu\nu}^{(i)}\big)\ ,
    \end{equation}
    where $g_{\mu\nu}^{(i)}$ is the metric in the Jordan frame and $g$ it's determinant. The metric in the Einstein frame is recovered through a conformal transformation $g_{\mu\nu}^{(i)}= e^{-\frac{2\, Q_{i}\, \phi}{c^2 \Mp}} \tilde{g}_{\mu\nu}$, with $Q_i$ the coupling of the chameleon field to each matter field, $\psi ^{(i)}$. As done in \textit{e.g.} \cite{Wilcox:2015kna,Butt:2024jes}, in the following, a single coupling constant $Q$\footnote{Note that the value $Q = 1/\sqrt{6}$, corresponds to the $f(\mathcal{R})$ models of gravity \cite{starobinsky2007disappearing, oyaizu2008nonlinear}.} for all matter components -- dark and baryonic -- will be assumed\footnote{While it is possible to extend the analysis to distinct field couplings to the chameleon field, it falls away from the objective of the manuscript. It will be explored in future investigations.}. $\Mp =1/\sqrt{8\pi G}$ is the reduced Planck mass, and $c$ is the speed of light. Non-relativistic matter have an energy density $\tilde{\rho}=\rho\, e^{\frac{Q\, \phi}{c^2 \Mp}}$.
    
    The potential $V(\phi)$ is a monotonic function of the scalar field, typically modelled as an inverse power-law:
    \begin{equation}\label{potential}
     V(\phi) = \lambda^{4+n} \phi^{-n} \ ,
    \end{equation}
    with $n\in \mathbb{N}$ and $\lambda$ represents an energy scale which can be set to the dark energy scale \cite{Tamosiunas_2022}.
    
    The scalar (chameleon) field equation resulting from the Euler-Lagrange equations of Eq.~\eqref{eq:lagrangian} in the quasi-static limit, is
    \begin{equation}\label{campophi}
     \nabla^2\phi = V'(\phi) + \frac{Q}{M_{\text{Pl}}c^2} \sum_j\rho_j e^{\frac{Q\, \phi}{M_{\text{Pl}}c^2}}\ ,
    \end{equation}
     where a prime denotes differentiation with respect to the scalar field, and the index $j$ runs over all involved non-relativistic matter species. It is worth mentioning that $\phi / M_{\text{Pl}} $ has the dimension of energy per unit mass, \textit{i.e.} plays the role of an additional gravitational potential. From the right-hand side of Eq.~\eqref{campophi} one can notice that the dynamics of $\phi$ are dictated by an effective potential $V_{\text{eff}}(\phi)$ that encapsulates both the potential $V(\phi)$ and the matter interaction feedback:
    \begin{equation} \label{eq:effective_potential}
     V_{\text{eff}}(\phi) \equiv V(\phi) + \sum_j\rho_j\, e^{\frac{Q\, \phi}{M_{\text{Pl}}c^2}} \ .
    \end{equation} 
    Current constraints on Chameleon gravity~\cite{Xang19,Desmond2020,Boumechta:2023qhd} limit the field background to be $\ll 1$ (in units of $c^2$). Thus, one can assume $Q\phi/(c^2 \Mp)\ll1$, and the effective potential can be approximated by $\Veff(\phi)\simeq V(\phi)+\rho_j\, \big( 1+\beta\phi/(c^2 \Mp)\big)$. The resulting equation of motion is, 
    \begin{equation} \label{eq:eq_of_motion_s}
     \nabla^{2}\phi=\frac{\beta}{c^2 \Mp}\sum_j\rho_j+V'(\phi)\ .
    \end{equation}
    The profile of the chameleon field and the resulting fifth-force is obtained by solving Eq.~\eqref{eq:eq_of_motion_s} with suitable boundary conditions once a model for the matter density distribution is chosen. In the following, we employ a semi-analytical approach to reconstruct the field profile~\cite{Terukina:2013eqa,Wilcox:2015kna,Pizzuti2024b}. This consists of dividing the spacetime into two regions: deep within the massive source, characterized by a field profile $\phi_{int}$, and towards the low-density outskirts, where the field is described by $\phi_{out}$. The semi-analytic solution is then established by linking the interior with the exterior solution.
    
    Deep inside the matter distribution, over-density region, the scalar field is suppressed and stays at the minimum of the effective potential, $\nabla^2\phi\approx 0$, effectively screening the fifth-force. From Eq.~\eqref{eq:eq_of_motion_s}, the resulting scalar field inside the matter distribution comes as
    \begin{equation} \label{campoint}
     \phi_{int} \approx \left( Q\, \frac{\rho_{tot}}{n \lambda^{4+n}M_{\text{Pl}}} \right)^{-\frac{1}{(n+1)}} \ ,
    \end{equation}
    where the total mass distribution is defined as $\rho_{tot} = \sum_j\rho_j$. On the other hand, the outer solution is obtained when the contribution of the scalar field potential, the first term on the right-hand side of Eq.~\eqref{eq:eq_of_motion_s}, is sub-dominant to the matter density and $\nabla^2\phi$. This describes the case where the chameleon field mediates a long-range fifth-force, the matter density is still large compared to the background, and the scalar field has not settled to the minimum of the effective potential. The equation of motion for the exterior chameleon field is thus given by:
    \begin{equation}\label{luna}
     \nabla^2\phi_{out} \approx Q \frac{\rho_{tot}}{M_{\text{Pl}}} \ .
    \end{equation}

    In the following, let us assume a spherically symmetric matter distribution, $\rho_{tot}\equiv\rho_{tot}(r)$, which is a valid assumption for galaxy clusters in dynamical equilibrium~\cite{Lagan2019,Biviano:2023oyf}; the effect of triaxiality of halos in the Chameleon mechanism is investigated in \cite{Tamosiunas_2022}. For spherically symmetric matter distributions Eq.~\eqref{luna} simplifies to:
    \begin{equation} \label{simmsferica}  
     \frac{1}{r^2} \frac{\text{d}}{\text{d}r} \left( r^2\frac{\text{d}\phi_{out}}{\text{d}r} \right) = Q\, \frac{\rho_{tot}(r)}{M_{\text{Pl}}} \ .
    \end{equation}
    A single integration gives:
    \begin{equation}\label{firstintegration}
     \frac{\text{d}\phi_{out}}{\text{d}r} r^2 = \frac{Q}{M_{\text{Pl}}} \bigintssss r^2 \rho_{tot}(r) \text{d}r + C \ ,
    \end{equation}
    with $C$ an integration constant. Notably, this expression shows that the fifth-force in the outer region is proportional to the total mass profile, given by the integral on the right-hand side of Eq.~\eqref{firstintegration}. The Poisson equation for the gravitational potential in this region can now be written as
    \begin{equation}\label{gradient}
     \frac{\text{d}\Phi}{\text{d}r} = \frac{GM(r)}{r^2} + \frac{Q}{M_\text{Pl}} \frac{\text{d}\phi}{\text{d}r} \ ,
    \end{equation}
    where the second term expresses the contribution of the fifth-force generated by the coupling between the chameleon field and the matter distribution. Eq.~\eqref{gradient} can be expressed in a Newtonian-like fashion by defining an \textit{effective mass},
    \begin{equation}
     M_{\text{eff}} = \frac{Q}{G} \frac{r^2}{M_\text{Pl}} \frac{\text{d}\phi}{\text{d}r} \ ,
    \end{equation}
    from which results the total dynamical mass:
    \begin{equation}\label{eq:massdyn}
     M_{\text{dyn}} = M_{\text{GR}} + M_{\text{eff}} \ .
    \end{equation}
    Note that, due to the null geodesic's invariance under conformal transformations, the structure of the Chameleon model produces lensing measurements that are only sensitive to the Newtonian part of the gravitational interaction, that is, the first term in Eq.~\eqref{gradient} \cite{Burrage:2017shh}. This means that lensing surveys can be used as a complementary probe for the prior of the mass profile.

    The scalar field can now be obtained as:
    \begin{equation}\label{campoout}
     \phi_{out} = \frac{Q}{M_{\text{Pl}}} \bigintssss \frac{1}{r^2} \bigg[\int (r')^2 \rho_{tot}(r') \text{d}r' + \Cs  \bigg] \text{d}r \ . 
    \end{equation}
    The integration constant $\Cs$ can then be determined by imposing continuity of \(\phi_{\text{int}}\) and \(\phi_{\text{out}}\), and its derivatives, at the matching screening radius $r = S$, which denotes the transition scale between the two regimes. The term "screening radius" is used since within a distance \(S\), the effects of modified gravity are negligible, while at distances greater than \(S\), the fifth-force becomes significant. In the inner core of the halo, the field is significantly suppressed compared to its ambient value, \(\phi_s \ll \phi_\infty\), making the interior solution negligible in comparison, \(\phi_{\text{int}} \approx 0\).

%
 \section{Multi-component chameleon solution}\label{sec:generalscreening}
%
    As already stated, in a galaxy cluster, the matter distribution, $\rho _j$, can be generally modelled as the sum of four main components (see \textit{e.g.} \cite{Sartoris2020,Laudato:2021mnm}): the Brightest Cluster Galaxy (BCG), $\rho_\text{BCG}$, which dominates the mass profile and the dynamics in the innermost region ($r\lesssim 0.05 \text{Mpc}$); the hot Intra-Cluster Medium (ICM), $\rho_\text{ICM}$; the mass density associated with the baryonic mass in galaxies (other than the BCG), $\rho_*$; and the diffuse dark matter component (CDM), $\rho_\text{CDM}$. If a cluster is dynamically relaxed, all the mass distributions should be in equilibrium with the total gravitational potential and exhibit a nearly spherical disposition.

    The total mass density, $\rho_\text{tot}$, is then
    \begin{equation}\label{eq:rhotot}
     \rho_\text{tot}(r) = \rho_\text{BCG}(r) + \rho_\text{ICM}(r) + \rho_{*}(r) + \rho_\text{CDM}(r)\ , 
    \end{equation}
    with each component modelled by the appropriated mass model. In particular, the BCG is, typically, a giant elliptical galaxy located close to the centre of the cluster's gravitational potential. The associated surface brightness distribution is well described by a de Vaucouleurs profile (\textit{e.g.} \cite{Biviano:2023oyf,Sartoris2020}). The de-projection of this profile is well approximated to a Jaffe profile~\cite{Jaffe83}, which exhibits a simple analytical form, and thus it will be used as a mass density model for the stellar distribution of the BCG, following~\cite{Sand2004}:
    \begin{equation}\label{eq:bcg}
     \rho_\text{BCG}(r) = \frac{\rho_b}{\left(\frac{r}{r_J}\right)^2\left(1+\frac{r}{r_J}\right)^2}\ ,
    \end{equation}
    with the characteristic density written in terms of the total stellar mass, $M_*$, as 
    \begin{equation}
     \rho_b = \frac{M_*}{4 \pi r_J^3}\ .
    \end{equation}
    As for the baryonic mass in galaxies, the profile is modelled by a NFW mass density profile which has been found to provide an adequate fit for the distribution of galaxies in clusters (\textit{e.g.}~\cite{Annunziatella14,barrena24}),
    \begin{equation}\label{eq:galnfw}
     \rho_*(r) = \frac{\rho_*}{\frac{r}{r_*}\left(1+\frac{r}{r_*}\right)^2}\ .
    \end{equation}
    The hot, Intra-Cluster gas of the ICM is modelled assuming an Isothermal $\beta-$profile~\cite{Ettori:2018tus,King1962}:
    \begin{equation}\label{eq:Iso}
     \rho_\text{ICM}(r)=\frac{\rho_{g}}{\Big[ \left(\frac{r}{r_{g}}\right)^{2}+1\Big]^{3\,\alpha/2}}\ ,
    \end{equation}
    where we set $\alpha = 1$ in order to guarantee convergence of the chameleon field profile to the background value at large radii (see \cite{Pizzuti2024b}). Observe that, while other profiles have been shown to better describe the ICM distribution in clusters (\textit{e.g.} \cite{Patej2015,Cao2016}), the relative simplicity of the Isothermal ansatz makes it a widely used model distribution for the ICM (\textit{e.g.} \cite{Laudato:2021mnm,Radiconi22}.

    Finally, for the diffuse dark matter distribution, a generalized NFW ($g$NFW hereafter) is assumed,
    \begin{equation} \label{eq:gNFW}
     \rho_\text{CDM}(r)=\frac{\rhos}{\left(\frac{r}{\rs}\right)^\gamma \left( 1+\frac{r}{\rs}\right)^{3-\gamma}}\ ,    
    \end{equation}
    where the inner slope is controlled by the additional real exponent $0< \gamma < 2$\footnote{The standard NFW case is recovered for $\gamma = 1$.}. Note that in the \textsc{MG-MAMPOSSt} analysis that will be presented in Section~\ref{sec:anal}, the parameter $r_{200}$ will be used instead of $r_\text{g}$. This is the radius of a sphere enclosing an over-density $200$ times the critical density of the universe at a given redshift. The corresponding mass is $M_{200}^{(CDM)} = 100\,H^2(z)(r_{200})^3/G$. The critical density can be obtained from $r_{200},\, r_\text{s}$ and $\gamma$ as
    \begin{equation*}
     \rho_\text{s} = \left(\frac{r_{200}}{\rs}\right)^{\gamma-3}\frac{M_{200}^{(CDM)}(3-\gamma)}{4\pi r_g ^3\, _2F_1\left(3-\gamma ,3-\gamma ;4-\gamma ;-\frac{r_{200}}{\rs}\right) }\ ,
    \end{equation*}
    where $ _2F_1(a,b,c,z)$ is the ordinary hypergeometric function.

\begin{figure}[h!]
     \centering
     \includegraphics[width=0.6\textwidth]{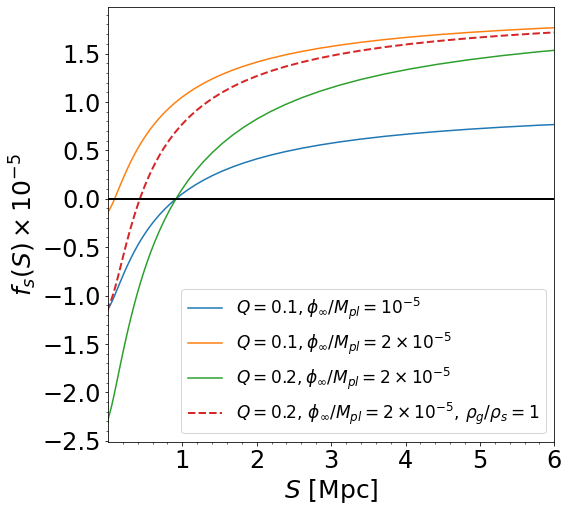}
     \caption{Solid lines: screening function $f_s(S)$ of Eq.~\eqref{eq:screenfun} for different values of the coupling and the value of the field at infinity (given in units of $c^2$. The density parameters are $\rhos = 10^{15}\,\text{M}_\odot/\text{Mpc}^3$, $\rs = 0.5 \,\text{Mpc}$, $\gamma = 0.5$, $\rho_* = 10^{13}\,\text{M}_\odot/\text{Mpc}^3$, $r_* = 0.3 \,\text{Mpc}$, $\rho_g = 10^{14}\,\text{M}_\odot/\text{Mpc}^3$, $r_g = 0.3 \,\text{Mpc}$, $\rho_b = 10^{15}\, \text{M}_\odot/\text{Mpc}^3$, $r_J = 0.03\, \text{Mpc}$. Red dashed line: setting $\rhos \equiv \rho_g = 5\times10^{14}\,\text{M}_\odot/\text{Mpc}^3$.}
     \label{fig:screenfunc}
    \end{figure}

    The solution of Eq.~\eqref{simmsferica}, assuming the expression of the density Eq.~\eqref{eq:rhotot}, gives the expression for the gradient of the field: 
    \begin{equation}
     \frac{\text{d}}{\text{dr}}\phi(r)=\left\{
        \begin{aligned}
         &\sim 0 & \text{if } r<S \ ,\\
         & \frac{\text{d}\phi_\text{out}}{\text{dr}}  & \text{if } r>S\ , 
        \end{aligned}
     \right.
    \end{equation}
    where
    \begin{equation} \label{eq:extdr}
        \begin{aligned}
         \frac{\text{d}\phi_\text{out}}{\text{dr}} = &\frac{\Cs}{r^2}+ \frac{Q}{\Mp r^2}\left\{ \frac{ \rhos r^{3-\gamma } \rs^{\gamma } \, _2F_1\left(3-\gamma ,3-\gamma ;4-\gamma ;-\frac{r}{\rs}\right)}{3-\gamma }-\frac{\rho_b r_J^4}{r+r_J}\right.\\
         & \left.+ \left[\rho_g r_g^3 \left(-\frac{r}{\sqrt{r^2+r_g^2}}-\log \left(\sqrt{r^2+r_g^2}-r\right)\right)+\rho_* r_*^3 \left(\frac{r_*}{r+r_*}+\log (r+r_*)\right)\right]\right\}\ .
        \end{aligned}
    \end{equation}
    By integrating Eq.~\eqref{eq:extdr} one obtains the exterior field profile ($r>S$),
    \begin{equation}
        \begin{aligned}
         \phi(r) = & -\frac{\Cs}{\Mp r}+\frac{Q}{\Mp r} \Bigg[-\frac{(-1)^{\gamma } \rhos \rs^3 \left(-\frac{r}{\rs}\right)^{3-\gamma } \, _2F_1\left(3-\gamma ,3-\gamma ;4-\gamma ;-\frac{r}{\rs}\right)}{3-\gamma }\\
         &-\rho_g r_g^3 \log \left(\sqrt{r^2+r_g^2}-r\right)
         + \rho_b r r_J^2 \log \left(\frac{r+r_J}{r}\right)
         +\rho_* r_*^3 \log (r+r_*)\Bigg] \\
         &+\frac{Q}{\Mp r} \left(\frac{\rhos \rs^2 r^{3-\gamma } (r+\rs)^{\gamma -2}}{\gamma -2}-\rho_b r_J^3+\rho_* r_*^3\right) +\frac{Q \rhos \rs^2}{(\gamma -2) \Mp}+\phi_\infty \ .
        \end{aligned}    
    \end{equation}
    The matching conditions at the screening radius, $\phi(S) = 0,\ \phi'(S) = 0$, provides an expression for the integration constant,
    \begin{equation}\label{eq:cs}
        \begin{aligned}
         \Cs = & \frac{Q}{\Mp}\left\{\frac{(-1)^{\gamma } \rhos \rs^3 \left(-\frac{S}{\rs}\right)^{3-\gamma } \, _2F_1\left(3-\gamma ,3-\gamma ;4-\gamma ;-\frac{S}{\rs}\right)}{3-\gamma }+\frac{\rho_b r_J^4}{S+r_J}\right.\\
         & \left. + \rho_g r_g^3 \left(\frac{S}{\sqrt{S^2+r_g^2}}+\log \left(\sqrt{S^2+r_g^2}-S\right)\right)-\rho_* r_*^3 \left(\frac{r_*}{S+r_*}+\log (S+r_*)\right)\right\}\ ,
        \end{aligned}
    \end{equation}
    with the screening radius $S$ given as the (real) numerical root of a "screening function": 
    \begin{equation}\label{eq:screenfun}
        \begin{aligned}
         f_s (S) \equiv\  & \frac{Q}{\Mp} \Bigg\{-\frac{\rho_g r_g^3}{\sqrt{S^2+r_g^2}}+\frac{\rho_b r_J^3}{S+r_J}+\frac{\rhos \rs^2}{\gamma -2} \left(1-\left(\frac{S}{S+\rs}\right)^{2-\gamma }\right)\\
         &-\frac{\rho_* r_*^3}{S+r_*}+\rho_b r_J^2 \log \left(\frac{S}{S+r_J}\right)\Bigg\}+\phi_\infty =0\ .
        \end{aligned}
    \end{equation}
    Note that $f_s$ is a monotonically increasing function in $S$; this implies that if $f_s > 0$ for $S\to 0$, then the cluster is unscreened. In Figure~\ref{fig:screenfunc}, the behaviour of the screening functions for a typical galaxy cluster is shown, varying the chameleon parameters (solid lines) and the gas and dark matter central densities (purple and red dashed lines, respectively). The intersection with the x-axis (marked as the black solid horizontal line) provides the value of the screening radius $S$. 
%
    \subsection{Comparison with a single NFW profile}\label{sec:compNFW}
%
    Before further analysis, it is worth comparing the new multi-component modelling of the chameleon field with the results obtained when a single NFW profile is used to model the total cluster mass. For a reliable comparison, let us consider a NFW model characterized by the same value of $r^\text{(tot)}_{200} = 2.15 \,\text{Mpc}$\footnote{Note the superscript $(\text{tot})$ is being used here to distinguish from $r_{200}$ of the dark matter mass density profile.} and $r^\text{(tot)}_{-2} = 0.80 \,\text{Mpc}$, where the latter is the radius at which the logarithmic slope of the density profiles is equal to $-2$. For the multi-component case $\rhos = 3.97\times 10^{14}\, \text{M}_\odot/\text{Mpc}^3$, $\rs = 0.87\,\text{Mpc}$, $\gamma = 0.7$ for the dark matter density, and $\rho_b = 10^{15}\, \text{M}_\odot/\text{Mpc}^3$, $r_J = 0.03\, \text{Mpc}$ for the BCG profile. These values roughly correspond to the best fit found from the analysis of \cite{Biviano:2023oyf}, where the kinematic analysis of cluster member galaxies and stellar velocity dispersion profile (VDP) of the BCG is applied to the data of the massive relaxed galaxy cluster MACS 1206 at $z=0.44$. This high-quality dataset, complemented with lensing information and X-ray data of the hot ICM, will be used to constrain the chameleon parameters in an upcoming work.

    As for the galaxies and gas profiles, assume $\rho_* = 3.55\times 10^{13}\,\text{M}_\odot/\text{Mpc}^3$, $r_* = 0.36 \,\text{Mpc}$, $\rho_g = 2.54\times 10^{14}\,\text{M}_\odot/\text{Mpc}^3$ and $r_g = 0.37 \,\text{Mpc}$, respectively. These were obtained by fitting the observed galaxies and gas mass profiles with Eqs.~\eqref{eq:galnfw} and \eqref{eq:Iso}\footnote{The mass profile of member galaxies has been obtained from \cite{Annunziatella14}. The gas mass was provided by A. Biviano from S. Ettori via private communication.}.

\begin{figure}[h!]
     \centering
     \includegraphics[width=1.1\textwidth]{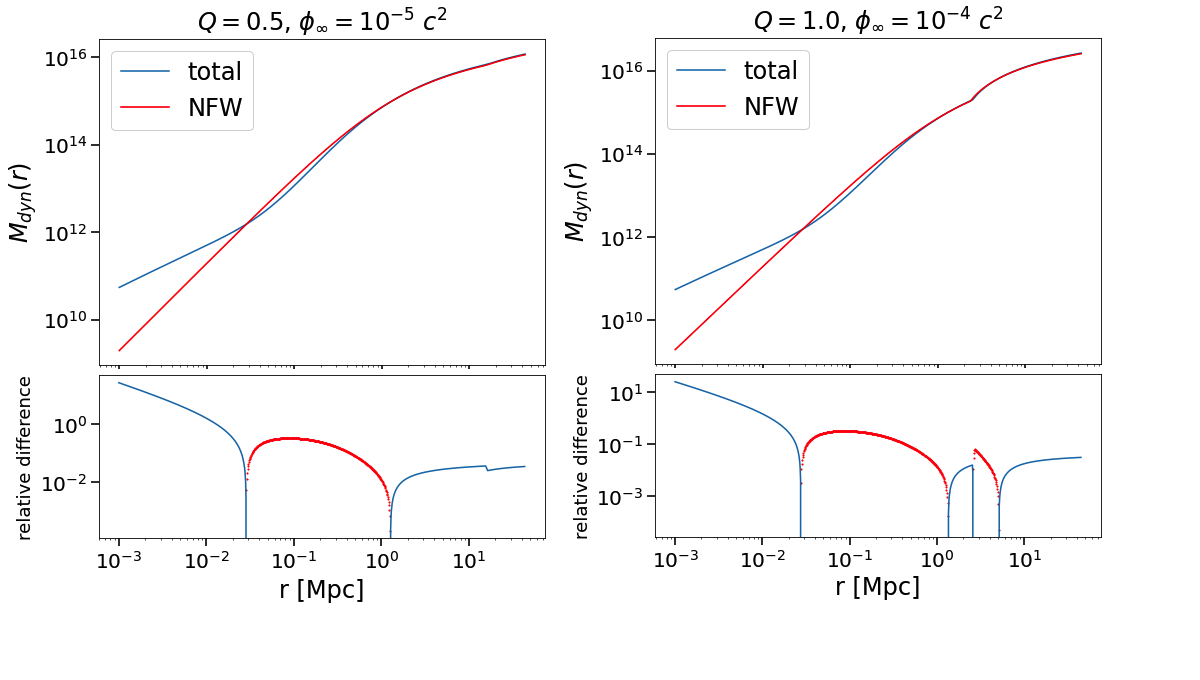}
     \caption{Top: total dynamical mass Eq.~\eqref{eq:massdyn}, for a NFW case (red) and the multi-component profile (blue), for two values of the coupling parameter and background field (left and right). Bottom: relative difference between the single and the multi-component profile. The parameters adopted for the mass components are $\rhos = 3.97\times 10^{14}$, $r_s = 0.87\,\text{Mpc}$, $\gamma = 0.7$, $\rho_* = 3.55\times 10^{13}\,\text{M}_\odot/\text{Mpc}^3$, $r_* = 0.36 \,\text{Mpc}$, $\rho_g = 2.54\times 10^{14}\,\text{M}_\odot/\text{Mpc}^3$, $r_g = 0.37 \,\text{Mpc}$,  $\rho_b = 10^{15}\,\text{M}_\odot/\text{Mpc}^3$, $r_J = 0.03\, \text{Mpc}$.}
     \label{fig:NFWcomparison}
    \end{figure}

    In the top panel of Figure~\ref{fig:NFWcomparison}, the total dynamical mass profiles, Eq.~\eqref{eq:massdyn}, for the multi-component case (blue) and the single NFW model (red) are shown. The respective relative difference $\left[M_\text{dyn}^{(multic)}- M_\text{dyn}^{(NFW)}\right]/M_\text{dyn}^{(NFW)}$ is presented in the lower plots. On average the two profiles exhibit a discrepancy $\sim 10\%$, which becomes larger at $r\lesssim 0.1\,\text{Mpc}$ (\textit{e.g.} $\lesssim 0.05\,r_{200}^{(\text{tot})}$), where the contribution of the BCG becomes relevant. The red dots in the lower plots represent the region where the difference becomes negative (\textit{i.e.} the NFW model predicts a higher total mass with respect to the multi-component case). The impact of the fifth-force can clearly be observed as a bump - at $r\sim 10 \,\text{Mpc}$ on the left and at $r\sim 2 \,\text{Mpc}$ on the right - in the lower plots. This discontinuity occurs due to the slightly smaller value of the screening radius in the NFW case than the one of the multi-component model.
%
 \section{Validation with numerical solutions}\label{sec:numerical}
%
    In order to validate the approach described in Section~\ref{sec:generalscreening}, let us compare the solutions obtained through the semi-analytical approach with the ones obtained via numerically solving Eq.~\eqref{eq:eq_of_motion_s}.

    The numerical solution of Eq.~\eqref{eq:eq_of_motion_s} was obtained through a $6^{th}$-order explicit Runge-Kutta integrator with the appropriate boundary conditions\footnote{Namely, $\nabla ^2 \phi \approx 0 $ at the centre of the mass distribution and $\text{d}\phi /\text{d}r =0$ at infinity.} imposed through a Newton-Rapshon root-finding method. An inner cutoff radius was imposed to avoid the divergence present at the centre of the mass density profiles of the BCG, baryonic mass in galaxies, and CDM models. To get the best fit for the semi-analytic approach, the value of the latter ranged between $1-10\%$ of the screening radius, $S$. The appropriate boundary condition at infinity was imposed by considering a numerically small value of the scalar field derivative, $\sim 10^{-8}$, at a scaled radius $x\equiv r/S$, several times larger than the size of the main mass distribution, $x_{max}\approx 10^3$. 

\begin{figure}
     \centering
     \includegraphics[width =1.\columnwidth]{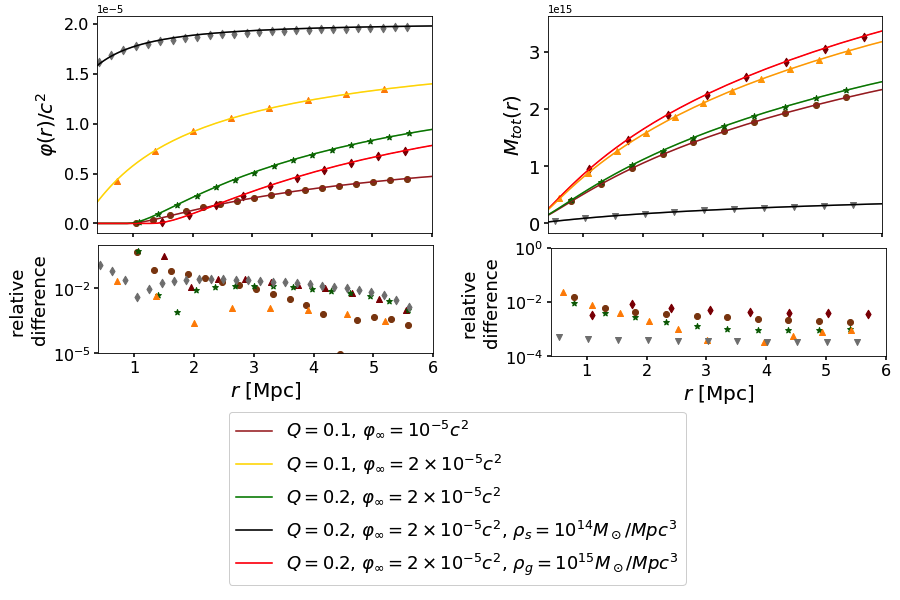}
     \caption{Left: Semi-analytic approximation for the radial field profile $\varphi(r) = \phi/\Mp$ (solid lines), compared with the numerical solution (points) for different values of the mass profile parameters. The bottom plot indicates the relative difference between the two. Right: the same for the total dynamical mass $M_\text{tot} = M+ M_\text{eff}$. The parameters adopted are $\rhos = 3.97\times 10^{14}$, $r_s = 0.87\,\text{Mpc}$, $\gamma = 0.7$, $\rho_* = 3.55\times 10^{13}\,\text{M}_\odot/\text{Mpc}^3$, $r_* = 0.36 \,\text{Mpc}$, $\rho_g = 2.54\times 10^{14}\,\text{M}_\odot/\text{Mpc}^3$, $r_g = 0.37 \,\text{Mpc}$,  $\rho_b = 10^{15}\, \text{M}_\odot/\text{Mpc}^3$ and $r_J = 0.03\, \text{Mpc}$.}   
     \label{fig:numerical}
    \end{figure}

    Comparative results between the semi-analytic (solid) and the full numerical solutions (points) can be seen in Figure~\ref{fig:numerical} for the multi-component mass density model assuming the mass configuration presented in Section~\ref{sec:compNFW} (Figure~\ref{fig:NFWcomparison}), and then varying the chameleon and density parameters.

    In Figure~\ref{fig:numerical}, both the scalar field profile (top left) and the total mass (top right) are represented, as well as the respective relative difference between the semi-analytically and the numerically-obtained solutions (bottom). 

    For the study of how well the semi-analytical solution describes the true behaviour of the scalar field, let us analyse the relative difference for the scalar field profile (bottom left). In the region inside the screening radius $S$, $x\in [0,1]$ -- where the strongest assumptions/approximations were made --, while the semi-analytical is set to zero, the numerical solutions are small but non-zero. This originates a large relative difference, which is accentuated at the transitional scaled radius $x_S$, where the numerics gain significant non-zero values before the semi-analytical (smoother transition from a negligible value). This behaviour is, however, cancelled as one goes away from the mass distribution ($x>1$) to the background configuration (no mass distribution, flat scalar field profile). At this point, the semi-analytical and the numerically obtained solutions coincide almost perfectly, with a maximum relative error of $10^{-3}$. 

    Therefore, besides some slight differences in the screening radius transition and asymptotic behaviour, the semi-analytic approximation describes the numerically obtained results with a high degree of accuracy, giving confidence in their use in more complex calculations that will proceed. 
%
 \section{Constraints on Chameleon gravity with kinematics and lensing analyses of galaxy clusters} \label{sec:anal}
%
    In what follows, we simulate the kinematic and lensing information of a galaxy cluster, as given by realistic current and upcoming surveys. The aim is to forecast the constraints on ($\phi_\infty,\, Q$) obtained when considering a multi-component mass reconstruction of the cluster. As done in \textit{e.g.} \cite{Wilcox:2015kna,Boumechta:2023qhd}, the analysis will be performed in terms of the scaled chameleon parameters  $\Qtwo=Q/(1+Q)$ and $\phitwo = 1- \exp\left[\phi/(10^4\,\Mp)\right]$, which range $[0,1]$.
    
    Following the procedure presented in \cite{Pizzuti2024b}, we generate a mock spherically symmetric, dynamically relaxed distribution of particles resembling an isolated galaxy cluster by using an updated version of the \textit{ClusterGEN} code~\cite{Pizzuti:2020tdl}. The features of this synthetic halo mimic the structure of MACS 1206, which will be the future target of the analysis presented here. In particular, we populated the halo up to $\sim 7\,r_{200}^\text{(tot)}$ -- where we set $r_{200}^\text{(tot)}=2.15\,\text{Mpc}$, see Section~\ref{sec:compNFW} -- considering 380 particles (\textit{aka} galaxies) within a sphere of $r_{200}^\text{(tot)}$; this way, we obtain 477 particles in a cylinder of (projected) radius $R=r^\text{(tot)}_{200}$, which is close to the number of galaxies in the reference sample of \cite{Biviano:2023oyf}. The particles are distributed according to a NFW number density profile $\nu(r) = \nu_\text{NFW}(r, r_\nu)$, where the scale radius is $r_\nu = 0.46 \,\text{Mpc}$, corresponding to the best fit quoted in \cite{Biviano:2023oyf} for the same sample. As for each component of the mass profile, we assume the values of the parameters displayed in Section~\ref{sec:compNFW}.
    
    To every galaxy (particle), at a 3-dimensional distance $r$ from the cluster centre, is assigned a velocity dispersion along the radial direction, $\sigma^2_r(r)$, given by the solution of the \textit{Jeans' equation}:
    \begin{equation}\label{eq:jeans}
     \frac{\text{d} (\nu \sigma_r^2)}{\text{d} r}+2\beta(r)\frac{\nu\sigma^2_r}{r}=-\nu(r)\frac{\text{d} \Phi}{\text{d} r}\ .
    \end{equation}
    Above, $\Phi$ is the total gravitational potential and $\beta(r) \equiv 1-(\sigma_{\theta}^2+\sigma^2_{\varphi})/2\sigma^2_r$ is the velocity anisotropy profile; $\sigma_{\theta}^2$ and $\sigma^2_{\varphi}$ are the velocity dispersion components along the tangential and azimuthal directions, respectively. Under the assumption of spherical symmetry, $\sigma_{\theta}^2=\sigma^2_{\varphi}$, and the anisotropy profile simplifies to $\beta(r) = 1 -\sigma_{\theta}^2/\sigma_{r}^2$. The mock halo is simulated assuming a generalization of the Tiret model~\cite{Tiret01} for $\beta(r)$ ($\beta_{gT}$ hereafter, see~\cite{Mamon2019}):
    \begin{equation} \label{eq:anisgT}
     \beta_{gT}(r) = \beta_0 +(\beta_\infty -\beta_0) \frac{r}{r+r_{\beta}}\ ,
    \end{equation}
    where $r_{\beta}$ is a scale radius that is assumed to be equal to $r_{-2} = (2-\gamma)\,\rs$ of the CDM density profile \cite{Biviano:2023oyf}; $\beta_0 = 0.5$, $\beta_\infty = 0.9$ are the values of the anisotropy at the centre and large radii, respectively, again chosen to mimic the radial anisotropy reconstructed for MACS 1206. Note that $\beta _{gT}$ can spawn from $-\infty$, for purely tangential orbits to $1$, for purely radial orbits. In what follows, let us define $\mathcal{A}_{0,\infty} = (1-\beta_{0,\infty})^{-1/2}$, which is strictly positive - $(0,\infty)$ - and $> (<)1$ for the radial (tangential) anisotropy, $=1$ for isotropic orbits.

    From the value of the anisotropy at $r$, the tangential (and azimuthal) velocity dispersion is given by
        \begin{equation}
            \sigma^2_\theta(r)\equiv \sigma^2_\varphi(r)  =\left[1-\beta(r)\right]\, \sigma^2_r(r)\ .
        \end{equation}
    Each component of the rest-frame velocities of particles can then be obtained by sampling over a Gaussian distribution $\mathcal{G}(\mu,\sigma_j)$, where $\mu = 0$ and $j =\{r,\,\theta,\,\varphi\}$.

    Along with the spatial (and momentum) distribution of the member galaxies in the synthetic cluster, a line-of-sight VDP  of the stars in the BCG has been simulated for six points in the projected radius $R$ from the centre of the galaxy -- assumed to coincide with the centre of the cluster. The line-of-sight VDP of the BCG will be indicated as $\sigma^2_\text{BCG}(R)$. As shown in \textit{e.g.}~\cite{Sartoris2020}, Eqs. (10) and (11), the VDP of the BCG depends on the gravitational potential sourced by all matter components. 
    
\begin{figure}
     \centering
     \includegraphics[width =1.\columnwidth]{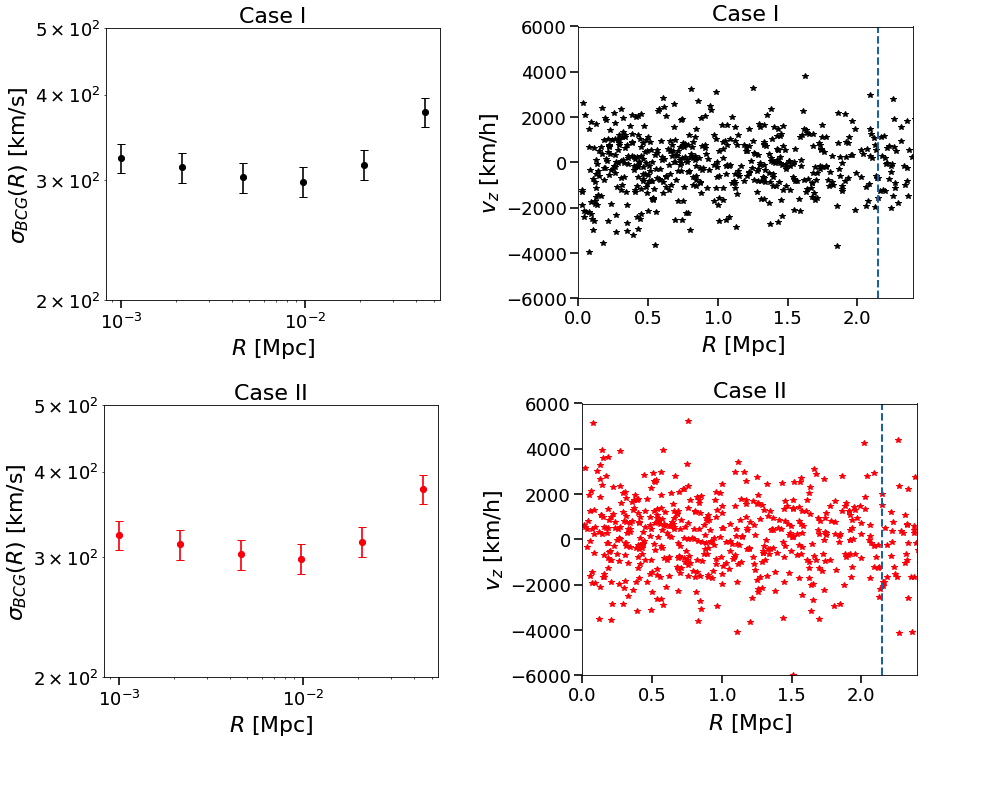}
     \caption{Mock data of the VDP along the line-of-sight for the BCG (left) and projected phase-space of member galaxies (right) for a multi-component modeled cluster generated assuming Newtonian gravity (top) and assuming a Chameleon universe with $\phi_\infty/\Mp=  10^{-4}\,c^2$ and $Q=1.0$ (bottom). The vertical blue dashed lines indicate the values of $r^\text{(tot)}_{200}= 2.15\,\text{Mpc}$.}    
     \label{fig:mockdata}
    \end{figure}

    In the left panels of Figure~\ref{fig:mockdata}, the simulated VDP is shown for the two sets of mock data considered in this analysis: Newtonian gravity, $\phi_\infty = Q = 0$ (hereafter \textbf{case I}); and a strong chameleon scenario $\phi_\infty/\Mp=  10^{-4}\,c^2$ and $Q=1.0$, as done in \cite{Pizzuti2024b} (hereafter \textbf{case II}). These values correspond to  $\phitwo = 0.63, \,\Qtwo=0.5$. Note that the VDP of the BCG is identical in both cases due to the large screening radius, $S = 2.63 \,\text{Mpc}$, that suppresses the dynamics of the fifth-force at the BCG scale (\textit{i.e.} the galaxy is completely screened).
    
    As for the uncertainties, we consider a $5\%$ error on each point, which is consistent with the uncertainties in the observational data used in \cite{Biviano:2023oyf}.
    To constrain the chameleon parameters from the mock cluster data, an upgraded version of the \textsc{MG-MAMPOSSt}\footnote{A previous version of the code is publicly available at \href{https://github.com/Pizzuti92/MG-MAMPOSSt}{https://github.com/Pizzuti92/MG-MAMPOSSt}.} code (see \cite{Pizzuti:2020tdl,Pizzuti:2022ynt}) is employed.

    Based on the \textsc{MAMPOSSt} method of \cite{Mamon01}, \textsc{MG-MAMPOSSt} performs a parametric reconstruction of the gravitational potential, $\Phi(r)$, the anisotropy $\beta(r)$ and the number density $\nu(r)$ profiles of spherically symmetric systems by solving the Jeans' equation in Newtonian gravity, or some general class of modified gravity. In particular, the code performs a Monte Carlo Markov-Chain (MCMC) sampling of the parameter space, with the projected phase-space (pps) of member galaxies as input data. This is the set of couples $(R^{(i)},v_z^{(i)})$, where $R^{(i)}$ is the projected position of a galaxy with respect to the centre of the cluster, and $v_z^{(i)}$ is the velocity measured along the line-of-sight (los). 

    In Figure~\ref{fig:mockdata} (right), the pps extracted from the synthetic clusters of case I (top) and case II (bottom) are shown. Note that, in general, the modified gravity scenario predicts a los velocity dispersion of member galaxies larger than in the Newtonian case.
    
    The \textsc{MG-MAMPOSSt} log-likelihood is given by
    \begin{equation}
     \mathcal{L}_\text{MAM}(\Theta) = \sum_i \text{ln}\,q(R_i,v_{z,i}| \Theta)\ ,
    \end{equation}
    with $q(R^{(i)},v_{z}^{(i)}| \Theta)$ the probability of finding a galaxy at the point $R^{(i)},v_{z}^{(i)}$. The sum runs over the particles in the pps, and $\Theta$ is the vector of parameters describing the model, in our case $\Theta \in \{r_{200},\,\rs,\,X_L,r_J,\phiinf,\,Q\}$, with $X_L = M_*/L_*$ the ratio between the stellar mass and the total luminosity of the BCG in a given band. Still following \cite{Biviano:2023oyf}, let us assume $L_*=4.9\times 10^{11}\,L_\odot$ in the $I$-band. Note that $\rho_g,\,r_g,\,\rho_*$ and $r_*$ are kept fixed to the best fit values quoted in Section~\ref{sec:compNFW}. It was further checked that variations of these parameters within the uncertainties does not provide considerable changes in the results of our analysis. 

    In addition to the pps, \textsc{MG-MAMPOSSt} is equipped with a module to fit the observed VDP of the BCG, such that the total kinematic log-likelihood is
    \begin{equation}
     \mathcal{L}_\text{kin}(\Theta) = \mathcal{L}_\text{MAM}(\Theta)-\frac{\chi_\text{BCG}(\Theta)}{2}\ ,
    \end{equation}
    with $\chi_\text{BCG}(\theta)$ the chi-square obtained by comparing the theoretical prediction of $\sigma^2_\text{BCG}(R_j)$ with the observed data points $\sigma^2_\text{obs,j}$ at $R_j$.

    As for the priors, we first consider broad, uninformative flat priors for all the free parameters in the analysis of both case I and case II, with the upper and lower bounds provided in the second and third columns of Table~\ref{tab:priors}. Note that for the BCG stellar mass-to-light ratio, we considered the bounds given by the two-$\sigma$ region from the analysis of \cite{Biviano:2023oyf}. 

\begin{figure}
     \centering
     \includegraphics[width =1.\columnwidth]{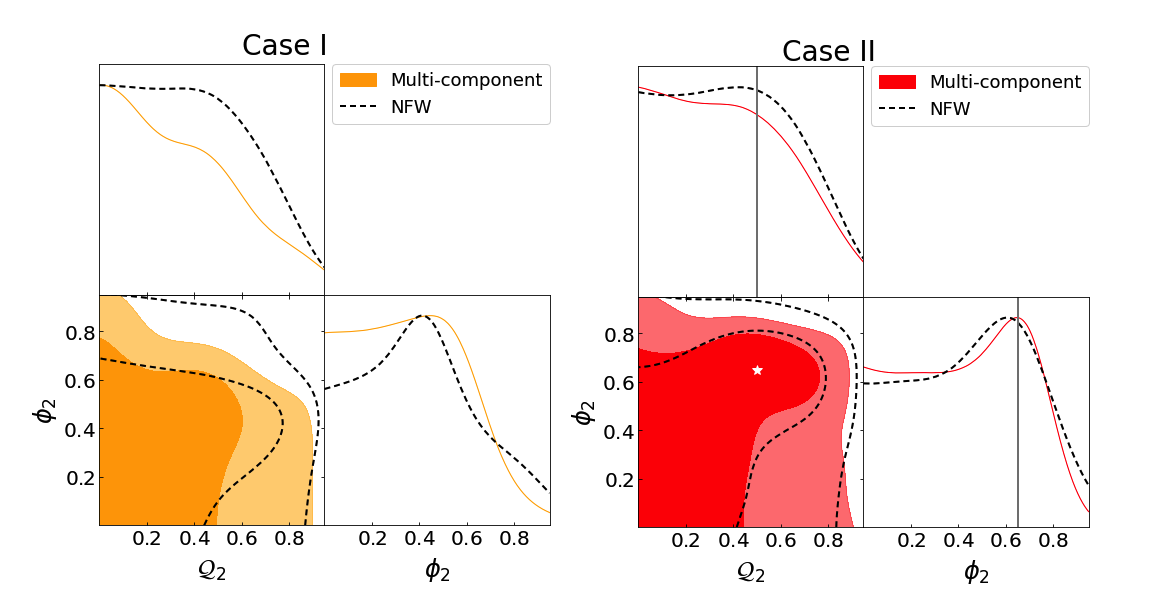}
      \caption{Solid lines: marginalized distributions of $\phitwo$ and $\Qtwo$ from the \textsc{MG-MAMPOSSt} analysis of the cluster in Newtonian gravity (left) and Chameleon gravity with $\Qtwo = 0.5$ and $\phitwo = 0.63$ (right). The white star and the vertical solid lines on the right plots indicate the true values of the chameleon parameters. The inner and outer shaded regions represent the one-$\sigma$ and two-$\sigma$ contours in the parameter space, respectively. Dashed line: distribution obtained when considering a single NFW mass profile to model the total mass distribution in the \textsc{MG-MAMPOSSt} fit.}    
     \label{fig:kin}
    \end{figure}

    The marginalized probability distributions of the chameleon parameters, resulting from a 110000 points-MCMC sampling of $ \mathcal{L}_\text{kin}$\footnote{The exploration of the parameter space has been meticulously performed using a Metropolis-Hastings algorithm, with the first 10000 points considered as burn-in phase.}, are shown in the left (case I) and right (case II) panels of Figure~\ref{fig:kin}. For comparison, the dashed lines represent the results obtained when the total mass of the cluster is modelled as a single NFW profile (in this case, no BCG data are considered). 

    As already shown in \cite{Pizzuti2024b} for a single mass model, the constraints for the case I agree with GR expectation ($\phitwo = \Qtwo = 0$) within one-$\sigma$. The contour of the single NFW slightly differs from that of the correct multi-component model; the only relevant changes that occur due to the addition of the BCG data to the kinematic analysis are for a region in the top right part of the parameter space ($\phitwo, \,\Qtwo$). Indeed, while it is true that the BCG VDP data is very powerful in constraining the central region of the cluster, including the slope of the mass profile -- as can be seen by the triangle plots in Appendix~\ref{app:triangle} -- at small $r$, the cluster is supposed to be screened, except for very large values of $\phitwo$ and $\Qtwo$ (\textit{i.e.} the top right part of the two-dimensional contour plots in Figure~\ref{fig:kin}).

    In case II, even if the true values of the chameleon parameters are sitting well inside the one-$\sigma$ region of the marginalized distribution, no evidence of departures from standard gravity can be claimed from the \textsc{MG-MAMPOSSt} analysis. This is not surprising, as already shown in previous works (\textit{e.g.} \cite{Pizzuti:2020tdl,Boumechta:2023qhd,Pizzuti2024b}), kinematic data in clusters cannot provide bounds in the chameleon space if no additional information is given. The reason is a quite strong degeneracy between the total cluster size (here represented by the freedom in $r_{200}$ of the CDM profile), $\Qtwo$ and $\phitwo$, as can be spotted in Figure~\ref{fig:noBCGcontour}. 

\begin{figure}
     \centering
     \includegraphics[width =1.\columnwidth]{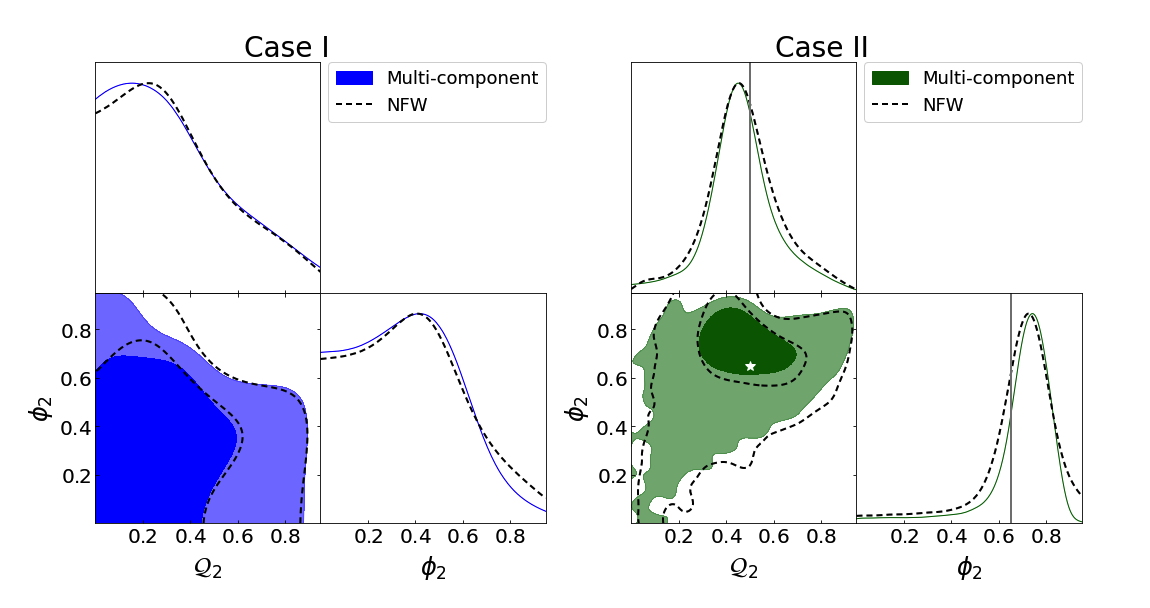}
     \caption{Same as Figure~\ref{fig:kin}, but including a Gaussian (lensing) prior on $r_{200}$ and $\rs$ of the CDM profile solid lines/filled areas: cluster where all the mass components are explicitly modelled. Dashed lines/contours: single NFW-modeled cluster.}    
     \label{fig:lens0}
    \end{figure}

    For this reason, we repeated the analysis by assuming a Gaussian prior on the CDM profile parameters $\mathcal{G}(r_{200},\,\rs)$ (note that the flat prior on the $g$NFW exponent $\gamma$ is kept unaltered), centered on the true values, which mimic the availability of a gravitational lensing survey. Following the prescription of \cite{Pizzuti2024b}, we have considered reliable uncertainties, given current lensing-like mass reconstruction \cite{Umetsu:2015baa}: $\sigma_{r_{200}} = 0.1\,r_{200}$ and $\sigma_{\rs} = 0.3\,r_\text{s}$, since the path of photons is unaffected by the fifth-force in Chameleon gravity. The gravitational lensing determinations of the total cluster's mass are thus sensitive only to the Newtonian part of the effective mass profile (\textit{e.g.} \cite{Schmidt_2009}). 
    \begin{table}[h!]
     \centering
     \caption{Priors values for the free parameters in the \textsc{MG-MAMPOSSt} analysis (kinematic only).\label{tab:priors}}
        \begin{tabular}{c|c|c}
         \toprule
         \textbf{Parameter} & \textbf{lower bound} & \textbf{upper bound}  \\
         \midrule
         $r_{200}$ & $0.5\,\text{Mpc}$ & $5.0\,\text{Mpc}$ \\
         $\rs$  & $0.05\,\text{Mpc}$ & $5.0\,\text{Mpc}$ \\
         $\gamma$  & 0 & 2\\
         $\mathcal{A}_\infty$ & 0.5 & 5.5 \\
         $\mathcal{A}_0$ & 0.5 & 5.5  \\
         $X_L$ & 4.20 & 4.74 \\
         $\phitwo$ & 0 & 1\\
         $\Qtwo$ & 0 & 1\\
         \bottomrule
        \end{tabular}
    \end{table}
    Figure~\ref{fig:lens0} shows the marginalized distributions for $\Qtwo$ and $\phitwo$ when the "lensing" prior is applied in the MCMC \textsc{MG-MAMPOSSt} run. Again, the dashed lines refer to a cluster with the same $r_{200}^{(tot)}$ and $ r_{-2}^{(tot)}$ modelled using a single NFW profile\footnote{In this case the Gaussian priors are centred on $r_{200}^{(tot)}$ and $ r_{-2}^{(tot)}$, with the same relative uncertainties.}. Interestingly, for case I, no relevant changes were obtained with the inclusion of additional information in the multi-component model. This is mainly due to the large values of $\Qtwo$ and $\phitwo$ being already excluded by the BCG VDP, as mentioned above. Note that the result of the NFW model + lensing is now almost identical to the multi-component case, confirming the effects of the Gaussian prior to cutting the top part of the parameter space.
    In a different fashion, the lensing information is required to distinguish strongly modified gravity scenario, where the combination of the \textsc{MG-MAMPOSSt} analysis plus the prior enhances the probability around the true values of the chameleon parameters (see right panel of Figure~\ref{fig:lens}). For this specific case, we obtain $\Qtwo =0.47^{+0.34 }_{-0.33}$ and $\phitwo =0.70^{+0.22 }_{-0.34}$ for the multi-component profile. 

    Finally, as shown in Figure~\ref{fig:lenscontourMG}, in clusters with strong MG signatures, the chameleon parameters exhibit a correlation with the anisotropy profile in the innermost regions of the cluster, which is absent in the standard scenario.
%
 \section{Conclusions}
%
   This manuscript presents a semi-analytical approach for describing the chameleon field in realistically modelled galaxy clusters. A particular focus was placed on the screening mechanism and the behaviour of the chameleon fifth-force when each matter component of a cluster was modelled by its individual best descriptor instead of a single model describing the total mass profile.

    The total galaxy cluster's mass density profile was first decomposed into four components: the BCG, the baryonic mass in galaxies, the ICM, and the CDM component. The BCG was described assuming a Jaffe profile and the ICM by an Isothermal beta model, while a NFW and gNFW profiles described the baryonic components in galaxies and the diffuse CDM, respectively. The semi-analytic expressions for the effective mass profile and the radial field profile were derived under spherical symmetry and tested against full numerical solutions.

     The semi-analytical approach was then implemented in the \textsc{MG-MAMPOSSt} code of \cite{Pizzuti:2022ynt}, which constrains modified gravity models at cluster scales with kinematics analyses of the member galaxies, assuming spherical symmetry and dynamical relaxation. As already discussed in \cite{Pizzuti2024b}, the advantage of the semi-analytic approximation relies on the reduced computational time with respect to the numerical computation while still yielding the same accuracy. This makes it suitable for statistical inference and fast exploration of the parameter space.

    By means of mock data -- based on the real massive galaxy cluster MACS 1206 at $z=0.44$ -- we forecast the obtainable bounds by a combination of pps data of galaxies, BCG (projected) VDP and additional lensing-like information on the CDM profile parameters $r_{200},\,\rs$. While for a GR cluster, the \textsc{MG-MAMPOSSt} analysis including BCG data is sufficient to break the degeneracy among mass profiles and chameleon parameters (excluding the large $(\Qtwo,\,\phitwo)$ part of the parameter space, contrary to what happens for a single mass model without the BCG data), in a modified gravity cluster the joint lensing+kinematic analysis is required in order to provide constraints on  $\Qtwo$ and $\phitwo$.
   
    The results from a multi-component-modelled cluster have been further compared with those obtained from a single NFW-modelled halo with the same total mass at $r = r^{\text{(tot)}}_{200}$. Interestingly, the stronger improvement on the constraints from the former comes from the possibility of including the BCG VDP data -- which can be used only upon an explicit model of the BCG stellar mass profile -- in the kinematic analysis.

    It is important to point out that the study performed here assumes clusters in dynamical equilibrium, and the mass components follow a spherical smooth distribution. As pointed out in \cite{Tamosiunas_2022}, large deviations from spherical symmetry may impact the effect of the fifth-force; moreover, in \cite{Pizzuti2020syst}, was discussed how the lack of dynamical relaxation is a relevant source of systematic effects for kinematics mass reconstruction in modified gravity. However, MACS 1206, which was the target of the analysis presented here, is a well-studied relaxed cluster with a nearly concentric mass distribution, as pointed out in several works (\textit{e.g.} \cite{Girardi2015,Biviano:2023oyf,ferrami2023dynamics}). 

    Along with kinematics and lensing, X-ray data can be used to further constrain cluster's mass profiles assuming hydrostatic equilibrium, both in GR (\cite{Ettori:2013tka}) and in MG (\textit{e.g.} \cite{Wilcox:2015kna,Boumechta:2023qhd}). As discussed in \cite{Pizzuti:2020tdl}, even if gas and galaxies "feel" the same gravitational potential, their physics is different, and so is the degeneracy among model parameters. Thus, in an upcoming work, the analysis presented here will be extended to include X-ray information. The constraining power of the joint kinematics, lensing and X-ray data will be explored and tested on a real cluster.

   Note that while the discussion here focused on the Chameleon model, the method can be easily extended to many alternative scenarios, such as other viable models in the Horndeski sector (\textit{e.g.}~\cite{Varieschi_2020}) or the more general Degenerate-Higher-Order Scalar-Tensor Theories (DHOST~\cite{Koyama:2015oma,Crisostomi:2019yfo,Dima_2021}), for which constraints at different scales have been recently derived in literature (\textit{e.g.}~\cite{Laudato:2021mnm,Haridasu:2021hzq,Saltas:2019ius,Hiramatsu_2022}). The formalism developed here, to simultaneously assess the inner- and outer-most regions of a galaxy cluster, is crucial to test scenarios where non-local interactions are present, for instance, non-minimal coupling of dark matter to gravity~\cite{Gandolfi:2021jai, Gandolfi:2023hwx, Zamani:2024oep} or fractional gravity~\cite{Benetti:2023imt}. Understanding the characteristic signature of different MG models on cluster's mass profile determinations is critical in the context of current and upcoming imaging and spectroscopic surveys at several wavelengths, both ground-based (\textit{e.g.} Vera Rubin Observatory\footnote{\url{https://www.lsst.org/about}}) and in space (Euclid ~\cite{EuclidI}, JWST~\cite{McElwain_2023}, Athena X-ray survey~\cite{Barret_2020} ), in order to provide a robust independent probe of possible alternatives to the $\Lambda$CDM scenario.
%
\section*{Acknowledgments}
%
This study is supported by the Italian Ministry for Research and University (MUR) under Grant 'Progetto Dipartimenti di Eccellenza 2023-2027' (BiCoQ). AMP is supported by the Czech Science Foundation (GACR) project PreCOG (Grant No. 24-10780S).


\appendix
\section[\appendixname~\thesection]{Marginalized distributions from the analysis of the mock clusters}
\label{app:triangle}
In the following, the marginalized distributions of all parameters in the \textsc{MG-MAMPOSSt} MCMC runs are displayed. In each plot, 2-dimensional inner and outer shaded regions correspond to one-$\sigma$ and two-$\sigma$ regions, respectively.
    \begin{figure}[h!]
     \centering
     \includegraphics[width =1.\columnwidth]{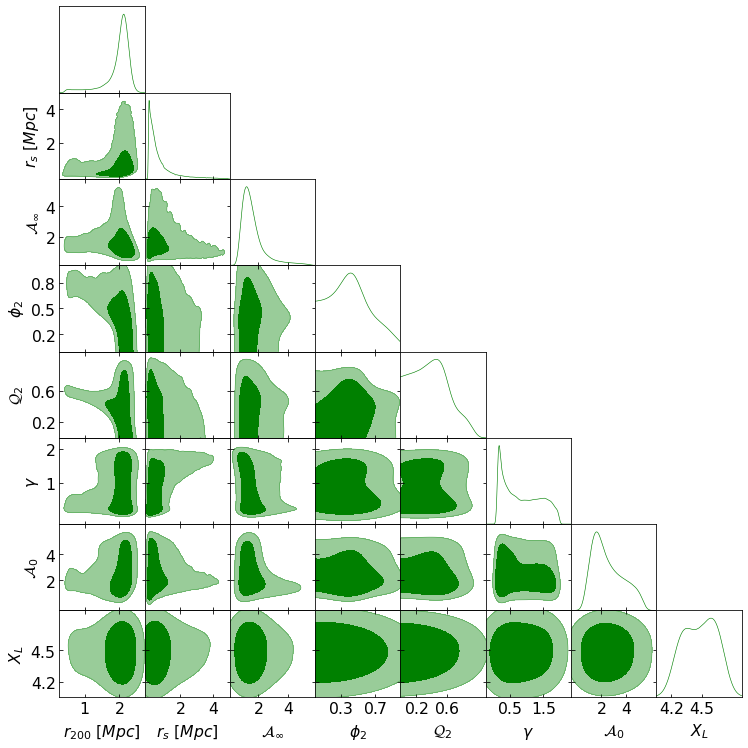}
     \caption{GR cluster, \textsc{MG-MAMPOSSt}, only pps of member galaxies.}    
     \label{fig:noBCGcontour}
    \end{figure}
    
    \begin{figure}[h!]
     \centering
     \includegraphics[width =1.\columnwidth]{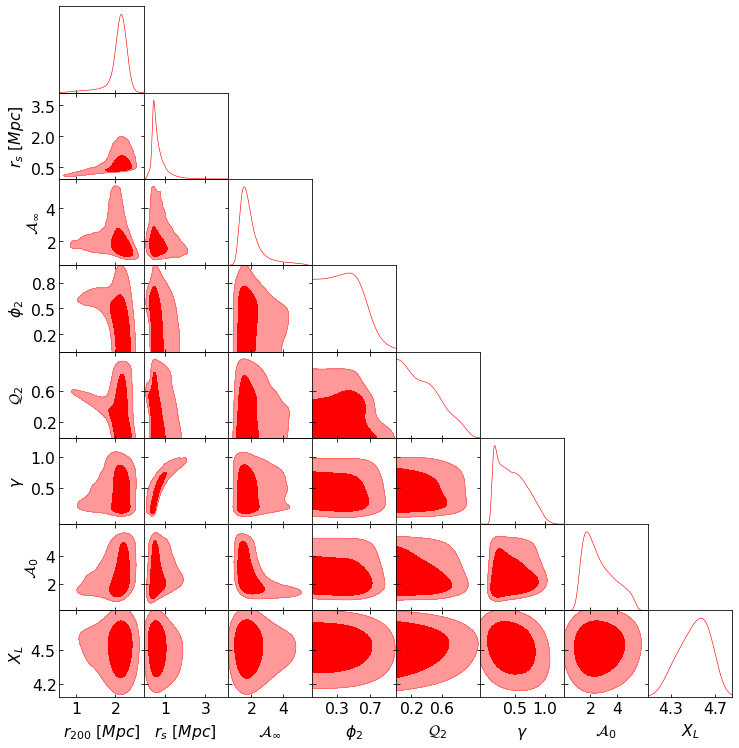}
     \caption{GR cluster, \textsc{MG-MAMPOSSt}, pps+BGC VDP.}    
     \label{fig:lens}
    \end{figure}
    \begin{figure}[h!]
     \centering
     \includegraphics[width =1.\columnwidth]{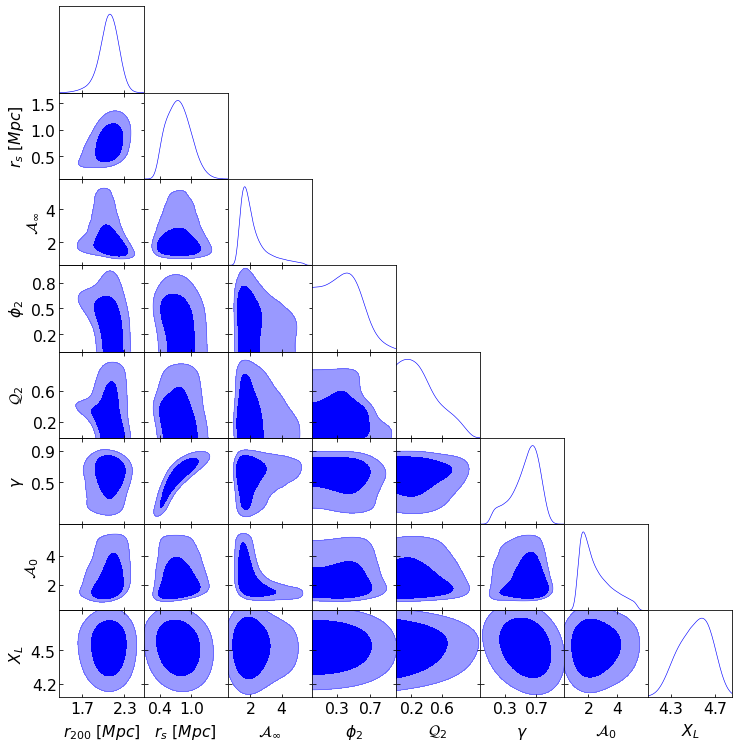}
     \caption{GR cluster, \textsc{MG-MAMPOSSt}, pps+BGC VDP + lensing prior.}    
     \label{fig:lenscontour}
    \end{figure}
    \begin{figure}[h!]
     \centering
     \includegraphics[width =1.\columnwidth]{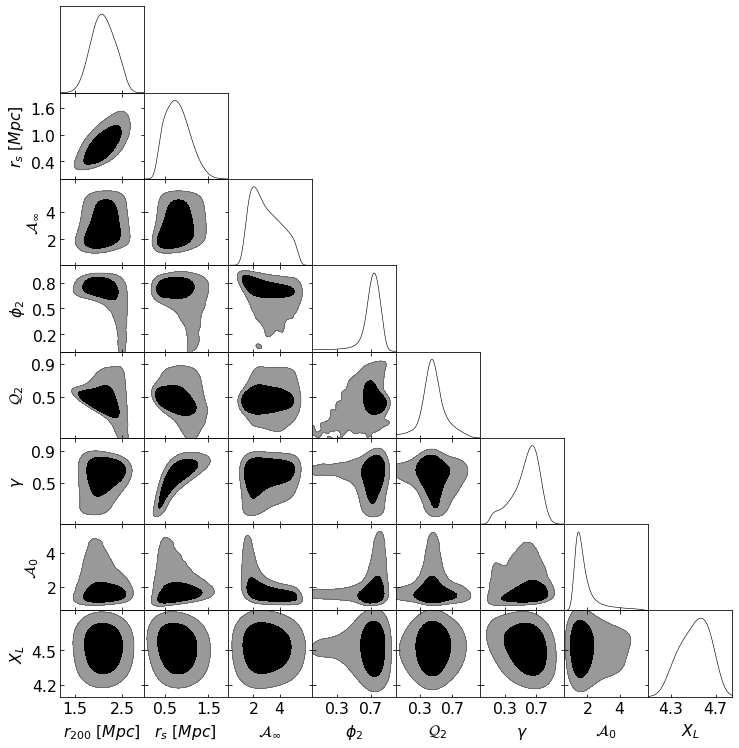}
     \caption{MG cluster, \textsc{MG-MAMPOSSt}, pps+BGC VDP + lensing prior.}    
     \label{fig:lenscontourMG}
    \end{figure}

\bibliographystyle{ieeetr} 
\bibliography{bibliography}

\end{document}